\documentclass[useAMS,usenatbib,referee]{biom}
\usepackage{amsmath}
\usepackage{natbib}
\usepackage{tabularx}
\long\def\/*#1*/{} 

\usepackage[figuresright]{rotating}
\usepackage{float}

\title[Utilizing stability criteria in choosing feature selection methods yields reproducible results in microbiome data]{Utilizing stability criteria in choosing feature selection methods yields reproducible results in microbiome data}

\author{Lingjing Jiang$^{1}$, Niina Haiminen$^{2}$,
        Anna-Paola Carrieri$^{3}$,
        Shi Huang$^{4,5}$, Yoshiki Vázquez-Baeza$^{4,5}$, \\
        {\bf Laxmi Parida$^{2}$,
       Ho-Cheol Kim$^{6}$, Austin D. Swafford$^{4}$, 
        Rob Knight$^{4,5,7,8}$, Loki Natarajan$^{1*}$}
        \email{lnatarajan@health.ucsd.edu}\\
	   $^{1}$Division of Biostatistics, University of California San Diego, La Jolla, CA 92093, USA \\
	   $^{2}$IBM T. J. Watson Research Center, Yorktown Heights, NY 10598, USA \\
	   $^{3}$IBM Research UK, The Hartree Center, Warrington, UK \\	   
	   $^{4}$Center for Microbiome Innovation, Jacobs School of Engineering, UC San Diego, La Jolla, CA 92093, USA \\
	   $^{5}$Department of Pediatrics, University of California San Diego, La Jolla, CA 92093, USA \\
	   $^{6}$Scalable Knowledge Intelligence, IBM Research-Almaden, San Jose, CA 95120, USA \\
	   $^{7}$Department of Computer Science and Engineering, University of California San Diego, La Jolla, CA 92093, USA\\
	   $^{8}$Department of Bioengineering, University of California San Diego, La Jolla, CA 92093, USA
	   }

\begin{document}



\pagerange{\pageref{firstpage}--\pageref{lastpage}} 


\label{firstpage}


\begin{abstract} 
Feature selection is indispensable in microbiome data analysis, but it can be particularly challenging as microbiome data sets are high-dimensional, underdetermined, sparse and compositional. Great efforts have recently been made on developing new methods for feature selection that handle the above data characteristics, but almost all methods were evaluated based on performance of model predictions. However, little attention has been paid to address a fundamental question: how appropriate are those evaluation criteria? Most feature selection methods often control the model fit, but the ability to identify meaningful subsets of features cannot be evaluated simply based on the prediction accuracy. If tiny changes to the training data would lead to large changes in the chosen feature subset, then many of the biological features that an algorithm has found are likely to be a data artifact rather than real biological signal. This crucial need of identifying relevant and reproducible features motivated the reproducibility evaluation criterion such as Stability, which quantifies how robust a method is to perturbations in the data. In our paper, we compare the performance of popular model prediction metric MSE and proposed reproducibility criterion Stability in evaluating four widely used feature selection methods in both simulations and experimental microbiome applications. We conclude that Stability is a preferred feature selection criterion over MSE because it better quantifies the reproducibility of the feature selection method.
\end{abstract}


\maketitle

\section{Introduction \label{section intro}}

Feature selection is indispensable for predicting  clinical or biological outcomes from microbiome data as researchers are often interested in identifying the most relevant microbial features associated with a given outcome. This task can be particularly challenging in microbiome analyses, as the datasets are typically high-dimensional, underdetermined (the number of features far exceeds the number of samples),  sparse (a large number of zeros are present), and compositional (the relative abundance of taxa in a sample sum to one). Current methodological research has been focusing on developing and identifying the best methods for feature selection that handle the above characteristics of microbiome data, however, methods are typically evaluated based on overall performance of model prediction, such as Mean Squared Error (MSE), R-squared or Area Under the Curve (AUC). While prediction accuracy is important, another possibly more biologically relevant criterion for choosing an optimal feature selection method is reproducibility, i.e. how reproducible are all discovered features in unseen (independent) samples? If a feature selection method is identifying true signals in a microbiome dataset, then we would expect those discovered features to be found in other similar datasets using the same method, indicating high reproducibility of the method. If a feature selection method yields a good model fit yet poor reproducibility, then its discovered features will mislead related biological interpretation. The notion of reproducibility for evaluating feature selection method seems intuitive and sensible, yet in reality we neither have access to multiple similar datasets to estimate reproducibility, nor have a well-defined mathematical formula to define reproducibility. The many available resampling techniques~\citep{efron1994introduction} enable us to utilize well-studied methods, for example bootstrapping, to create replicates of real microbiome datasets for estimating reproducibility. Moreover, given the burgeoning research in reproducibility estimation in the field of computer science~\citep{kalousis2005stability, kalousis2007stability, nogueira2018quantifying}, we can borrow their concept of Stability to approximate the reproducibility of feature selection methods in microbiome data analysis.  

In this paper, we investigate the performance of a popular model prediction metric MSE and the proposed feature selection criterion Stability in evaluating four widely used feature selection methods in microbiome analysis (lasso, elastic net, random forests and compositional lasso)~\citep{tibshirani1996regression, zou2005regularization, breiman2001random, lin2014variable}. We evaluate both extensive simulations and experimental microbiome applications, with a focus of feature selection analysis in the context of continuous outcomes. We find that Stability is a superior feature selection criterion to MSE as it is more reliable in discovering true and biologically meaningful signals. We thus suggest microbiome researchers use a reproducibility criterion such as Stability instead of a model prediction performance metric such as MSE for feature selection in microbiome data analysis. 

\section{Methods\label{method}}
\subsection{Estimation of stability} 

The Stability of a feature selection method was defined as the robustness of the feature preferences it produces to differences in training sets drawn from the same generating distribution~\citep{kalousis2005stability}. If the subsets of chosen features are nearly static with respect to data changes, then this feature selection method is a \textit{stable} procedure. Conversely, if small changes to the data result in significantly different feature subsets, then this method is considered \textit{unstable}, and we should not trust the output as reflective of the true underlying structure influencing the outcome being predicted. In biomedical fields, this is a proxy for reproducible research, in the latter case indicating that the biological features the method has found are likely to be a data artifact, not a real clinical signal worth pursuing with further resources~\citep{lee2013robustness}. \citet{goh2016evaluating} recommend augmenting statistical feature selection methods with concurrent analysis on stability and reproducibility to improve the quality of selected features prior to experimental validation~\citep{sze2016looking, duvallet2017meta}. 

While the intuition behind the concept of stability is simple, there is to date no single agreed-upon measure for precisely quantifying stability. Up to now, there have been at least~16 different measures proposed to quantify the stability of feature selection algorithms in the field of computer science~\citep{nogueira2017stability}. Given the variety of stability measures published, it is sensible to ask: which stability measure is most valid in the context of microbiome research? A multiplicity of methods for stability assessment may lead to publication bias in that researchers may be drawn toward the metric that extracts their hypothesized features or that reports their feature selection algorithm as more stable~\citep{boulesteix2009stability}. Under the perspective that a useful measure should obey certain properties that are desirable in the domain of application, and provide capabilities that other measures do not, Nogueira and Brown aggregated and generalized the requirements of the literature into a set of five properties~\citep{nogueira2017stability}. The first property requires the stability estimator to be fully defined for any collection of feature subsets, thus allowing a feature selection algorithm to return a varying number of features. The second property requires the stability estimator to be a strictly decreasing function of the average variance of the selection of each feature. The third property requires the stability estimator to be bounded by constants not dependent on the overall number of features or the number of features selected. The fourth property states that a stability estimator should achieve its maximum if and only if all chosen feature sets are identical. The fifth property requires that under the null model of feature selection, where we independently draw feature subsets at random, the expected value of a stability estimator should be constant. These five properties are desirable in any reasonable feature selection scenario, and are critical for useful comparison and interpretation of stability values. Among all the existing measures, only Nogueira’s stability measure (defined below) satisfies all five properties, thus we adopted this measure in the current work.

We assume a data set of $n$ samples $\{x_i,y_i\}_{i=1}^n$ where each $x_i$ is a $p$-dimensional feature vector and $y_i$ is the associated biological outcome. The task of feature selection is to identify a feature subset, of size $k<p$, that conveys the maximum information about the outcome $y$. An ideal  approach to measure stability is to first take $M$ data sets drawn randomly from the same underlying population, to apply feature selection to each data set, and then to measure the variability in the $M$ feature sets obtained. The collection of the $M$ feature sets can be represented as a binary matrix $Z$ of size $M \times p$, where a row represents a feature set (for a particular data set) and a column represents the selection of a given feature over the $M$ data sets as follows 
\begin{equation*}
Z = 
\begin{pmatrix}
Z_{1,1} & \cdots & Z_{1,p} \\
\vdots  & \ddots & \vdots  \\
Z_{M,1} & \cdots & Z_{M,p} 
\end{pmatrix}
\end{equation*}

Let $Z_{.f}$ denote the $f^{th}$ column of the binary matrix $Z$, indicating the selection of the $f^{th}$ feature among the $M$ data sets. Then $Z_{.f} \sim Bernoulli (p_f)$, where $\hat p_f =  \frac{1}{M}  \sum_{i=1}^M Z_{i,f}$  as the observed selection probability of the $f^{th}$ feature.  Nogueira defined the stability estimator as 

\begin{equation}
\hat \Phi(Z) = 1-  \frac{\frac{1}{p} \sum_{f=1}^p \sigma_f^2}{E [\frac{1}{p} \sum_{f=1}^p \sigma_f^2 |H_0 ]} 
= 1-\frac{\frac{1}{p} \sum_{f=1}^p \sigma_f^2 }{\frac{\bar k}{p}  (1- \frac{\bar k}{p})}
\end{equation}
where $\sigma_f^2=  \frac{M}{M-1} \hat p_f (1-\hat p_f)$ is the unbiased sample variance of the selection of the $f^{th}$ feature, $H_0$ denotes the null model of feature selection (i.e. feature subsets are drawn independently at random), and $\bar k =  \frac{1}{M}  \sum_{i=1}^M \sum_{f=1}^p Z_{i,f}$  is the average number of selected features over the $M$ data sets. 

In practice, we usually only have one data sample (not $M$), so a typical approach to measure stability is to first take $M$ bootstrap samples of the provided data set, and apply the procedure described in the previous paragraph. Other data sampling techniques can be used as well, but due to the well understood properties and familiarity of bootstrap to the community, we adopt the bootstrap approach. 

\subsection{Four selected feature selection methods }
Lasso, elastic net, compositional lasso and random forests were chosen as benchmarked feature selection methods in this paper due to their wide application in microbiome community~\citep{knights2011supervised}. Lasso is a penalized least squares method imposing an $L_1$-penalty on the regression coefficients~\citep{tibshirani1996regression}. Owing to the nature of the $L_1$-penalty, lasso does both continuous shrinkage and automatic variable selection simultaneously. One limitation of lasso is that if there is a group of variables among which the pairwise correlations are very high, then lasso tends to select one variable from the group and ignore the others. Elastic net is a generalization of lasso, imposing a convex combination of the $L_1$ and $L_2$ penalties, thus allowing elastic net to select groups of correlated variables when predictors are highly
correlated~\citep{zou2005regularization}. Compositional lasso is an extension of lasso to compositional data analysis~\citep{lin2014variable}, and it is one of the most highly cited compositional feature selection methods in microbiome analysis~\citep{kurtz2015sparse, li2015microbiome, shi2016regression, silverman2017phylogenetic}. Compositional lasso, or the sparse linear log-contrast model, considers variable selection via $L_1$ regularization. The log-contrast regression model expresses the continuous outcome of interest as a linear combination of the log-transformed compositions subject to a zero-sum constraint on the regression vector, which leads to the intuitive interpretation of the response as a linear combination of log-ratios of the original composition. Suppose an $n \times p$ matrix $X$ consists of $n$ samples of the composition of a mixture with $p$ components, and suppose $Y$ is a response variable depending on $X$. The nature of composition makes each row of $X$ lie in a $(p-1)$-dimensional positive simplex $S^{p-1}=\{(x_1,…,x_p ): x_j>0,j=1,…,p \text{ and } \sum_{j=1}^p x_j =1 \}$. This compositional lasso model is then expressed as
\begin{equation}
    y=Z \beta + \epsilon, \sum_{j=1}^p \beta_j =0
\end{equation}
where $Z=(z_1,…,z_p )=(logx_{ij})$ is the $n \times p$ design matrix, and $\beta= (\beta_1,…,\beta_p)^T$ is the $p$-vector of regression coefficients. Applying the $L_1$ regularization approach to this model is then
\begin{equation}
\hat \beta = argmin (\frac{1}{2n} ||y - z\beta||_2^2 + \lambda ||\beta||_1), \text{ subject to } \sum_{j=1}^p \beta_j = 0
\end{equation}
where $\beta = (\beta_1,…, \beta_p)^T, \lambda>0$ is a regularization parameter, and $||.||_2$ and $||.||_1$ denote the $L_2$ and $L_1$ norms, respectively. 

Random forests is regarded as one of the most effective machine learning techniques for feature selection in microbiome analysis \citep{belk2018microbiome, liu2017experimental, namkung2020machine, santo2019clustering, statnikov2013comprehensive}. Random forests is a combination of tree predictors such that each tree depends on the values of a random vector sampled independently and with the same distribution for all trees in the forest \citep{breiman2001random}. Since random forests do not select features but only assign importance scores to features, we choose features from random forests using Altmann’s permutation test \citep{altmann2010permutation}, where the response variable is randomly permuted $S$ times to construct new random forests and new importance scores computed. The $S$ importance scores are then used to compute the p-value for the feature, which is derived by computing the fraction of the $S$ importance scores that are greater than the original importance score. 

\subsection{Simulation settings}

We compared the performance of the popular model prediction metric MSE and the proposed criterion Stability in evaluating four widely used feature selection methods for different data scenarios. We simulated features with Independent, Toeplitz and Block correlation structures for datasets with the number of samples and features in all possible combinations of $(50, 100, 500, 1000)$, resulting in the ratio of $p$ (number of features) over $n$ (number of samples) ranging from 0.05 to 20. Our simulated compositional microbiome data are an extension of the simulation settings from \citet{lin2014variable} as follows: 
\begin{enumerate}
    \item Generate an $n \times p$ data matrix $W=(w_{ij})$ from a multivariate normal distribution $N_p(\theta,\Sigma)$. To reflect the fact the components of a composition in metagenomic data often differ by orders of magnitude, let $\theta = (\theta_j)$ with $\theta_j =log(0.5p)$ for $j=1,…,5$ and $\theta_j=0$ otherwise. To describe different types of correlations among the components, we generated three general correlation structures: Independent design where covariates are independent from each other, Toeplitz design where $\Sigma =(\rho^ {|i-j|})$ with $\rho=0.1,0.3,0.5,0.7,0.9$, and Block design with 5 blocks, where the intra-block correlations are 0.1, 0.3, 0.5, 0.7, 0.9, and the inter-block correlation is 0.09.
    \item Obtain the covariate matrix $X=(x_{ij})$ by the transformation $x_{ij} = \frac{exp(w_{ij})}{\sum_{k=1}^p exp(w_{ik})}$, and the $n \times p$ log-ratio matrix $z=log(X)$, which follows a logistic normal distribution \citep{aitchison1982statistical}. 
    \item Generate the responses $y$ according to the model $y=Z \beta^*+ \epsilon$, $\sum_{j=1}^p \beta_j^* =0$, where $\epsilon \sim N(0, 0.5^2)$, and $\beta^*=(1,-0.8,0.6,0,0,-1.5,-0.5,1.2,0,…,0)^T$, indicating that only 6 features are real signals. 
    \item Repeat steps 1-3 for 100 times to obtain 100 simulated datasets for each simulation setting, and apply the desired feature selection algorithm with 10-fold cross-validation on the 100 simulated datasets. Specifically, each simulated dataset is separated into training and test sets in the ratio of 8 : 2, 10-fold cross-validation is applied to the training set ($80\%$ of the data) for parameter tuning and variable selection, and then model prediction (i.e. MSE) is evaluated on the test set ($20\%$ of the data). Hence, stability is measured according to Nogueira’s definition based on the 100 subsets of selected features. Average MSE is calculated as the mean of the MSEs across the 100 simulated datasets, and the average false positive or false negative rate denotes the mean of the false positive or false negative rates across the 100 simulated datasets.  
\end{enumerate}

In summary, a total of 176 simulation scenarios were generated, with 16 for Independent design, 80 for Toeplitz or Block design, and 100 replicated datasets were simulated for each simulation setting, resulting in 17,600 simulated datasets in total. 

\section{Simulation results}
\begin{figure}
 \centerline{\includegraphics[width=5in]{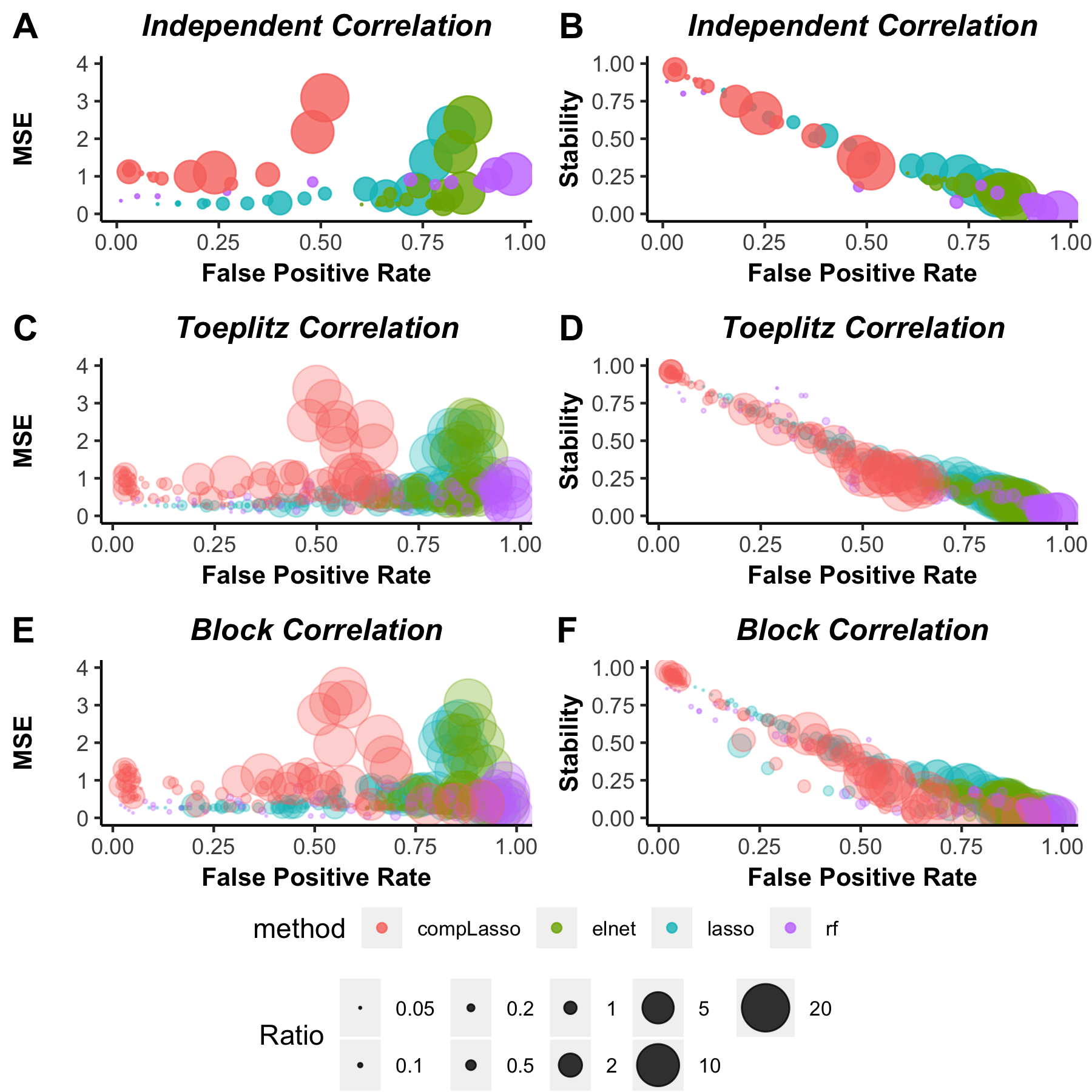}}
\caption{Comparing the relationship between MSE and False Positive Rate vs. Stability and False Positive Rate in three correlation structures. Colored dots represent values from different feature selection methods: compositional lasso (red), elastic net (green), lasso (blue) and random forests (purple). Size of dots indicate features-to-sample size ratio  $p/n$.}
\label{f:fig1}
\end{figure}

Given that the true numbers of false positive and false negative features are known in simulations, we can utilize their relationships with MSE and Stability to compare the reliability of MSE and Stability in evaluating feature selection methods. In theory, we would expect to see a positive correlation between MSE and false positive rate or false negative rate, while a negative correlation between Stability and false positive or false negative rates. This is because when the real signals are harder to select (i.e. increasing false positive or false negative rates), a feature selection method would perform worse (i.e. increasing MSE or decreasing Stability). The first column in Figure~\ref{f:fig1} shows the relationship between MSE and false positive rate in three correlation designs, and the second column in Figure~\ref{f:fig1} shows the relationship between Stability and false positive rate. In contrast to the random pattern in MSE vs. false positive rate (Figure~\ref{f:fig1} A-C-E), where drastic increase in false positive rate could lead to little change in MSE (e.g. random forests), or big drop in MSE corresponds to little change in false positive rate (e.g. elastic net), we see a clear negative correlation pattern between Stability and false positive rate (Figure~\ref{f:fig1} B-D-F). Regarding false negative rate, we also observe a random pattern in MSE and a meaningful negative correlation relationship in Stability (Supplementary Figure 1). These results suggest that Stability is a more reliable evaluation criterion than MSE due to its closer reflection of the ground truth in the simulations (i.e. false positive \& false negative rates), and this is true irrespective of feature selection method used, features-to-sample size ratio ($p/n$) or correlation structure among the features. 

Using the more reliable criterion Stability, we now investigate the best feature selection method in different simulation scenarios. Based on Stability, compositional lasso has the highest stability in “easier” correlation settings (Toeplitz 0.1 – 0.7 in Supplementary Figure 2, represented by Toeplitz 0.5 in Figure \ref{f:fig2} A due to their similar results; Block 0.9-0.3 in Supplementary Figure 3, represented by Block 0.5 in Figure \ref{f:fig2} C) for all combinations of $n$ (number of samples) and $p$ (number of features). Across all ``easier'' correlation scenarios, compositional lasso has an average stability of 0.76 with its minimum at 0.21 and its maximum close to 1 (0.97), while the 2nd best method Lasso has an average stability of only 0.44 with the range from 0.09 to 0.89, and the average stabilities of random forests and Elastic Net hit as low as 0.24 and 0.17 respectively. In ``extreme” correlation settings (Toeplitz 0.9 in Figure \ref{f:fig2} B or Block 0.1 in Figure \ref{f:fig2} D), compositional lasso no longer maintains the highest stability across all scenarios, but it still has the highest average stability of 0.42 in Toeplitz 0.9 (surpassing the 2nd best Lasso by 0.09), and the second highest average stability in Block 0.1 (only 0.03 lower than the winner Lasso). Regarding specific scenarios in “extreme” correlation settings, compositional lasso, lasso or random forests can be the best in different combinations of $p$ and $n$. For example, in both Toeplitz 0.9 and Block 0.1, with small $p$ (when $p$ = 50 or 100), random forests has highest stability ($\geq 0.8$) when $n$ is largest ($n=1000$), but Lasso or compositional lasso surpasses random forest when n is smaller than 1000, although all methods have poor stability ($\leq 0.4$) when $n \leq 100$. This indicates that best feature selection method based on Stability depends on the correlation structure among features, the number of samples and the number of features in each particular dataset; thus there is no single omnibus best, i.e., most stable,  feature selection method. 

\begin{figure}
 \centerline{\includegraphics[width=5in]{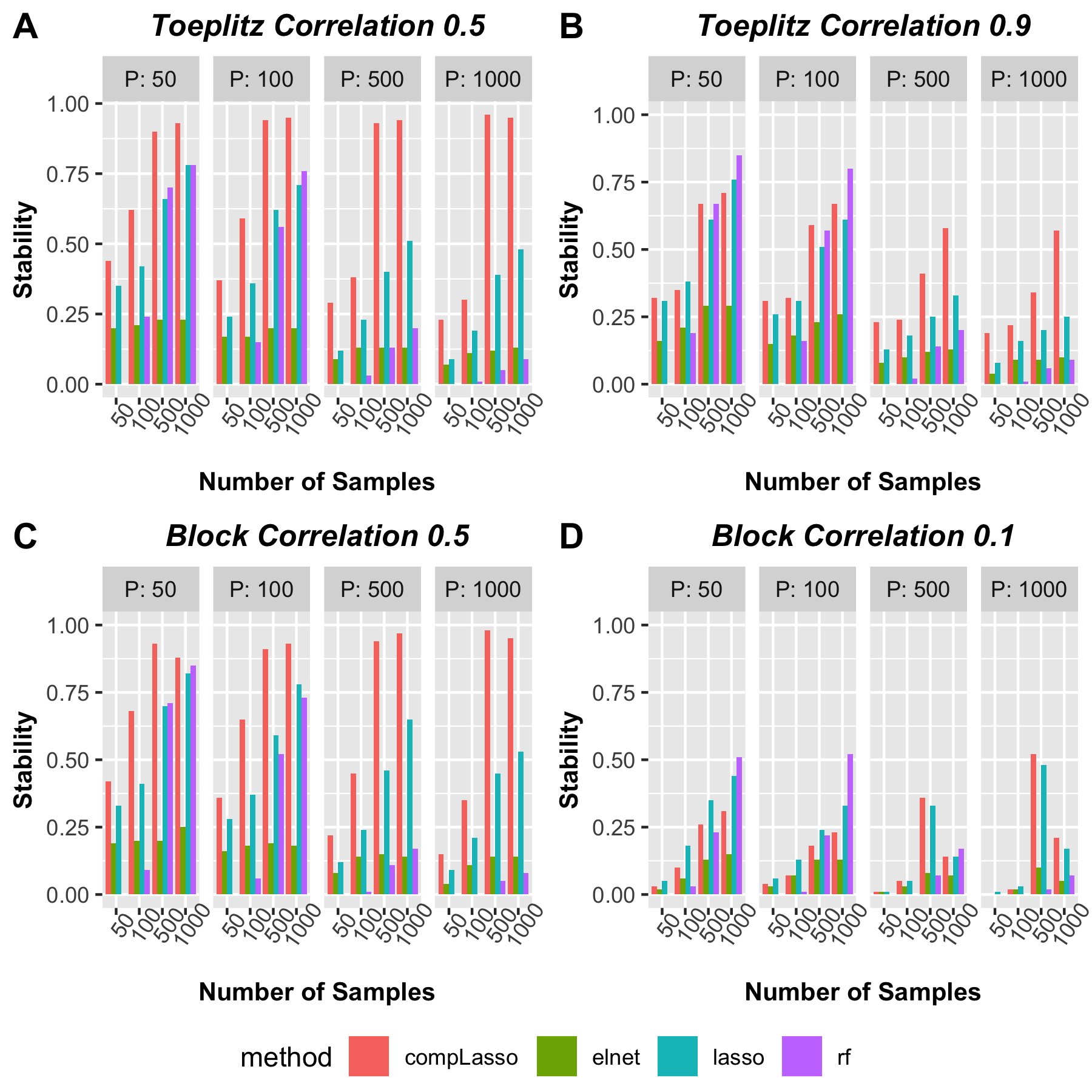}}
\caption{Method comparisons based on Stability in representative correlation structures. Colored bars represent Stability values corresponding to specific number of samples (x-axis) and number of features ($p$) for different feature selection methods: compositional lasso (red), elastic net (green), lasso (blue) and random forests (purple). Note that Toeplitz 0.1-0.7 has similar results as Toeplitz 0.5 (see Supplementary Figure 2), and Block 0.9-0.3 has similar results as Block 0.5 (see Supplementary Figure 3). Moreover, Stability equals to zero when no features were selected by methods (e.g. random forests chooses nothing when the number of samples equals 50).}
\label{f:fig2}
\end{figure}

\begin{figure}
 \centerline{\includegraphics[width=5in]{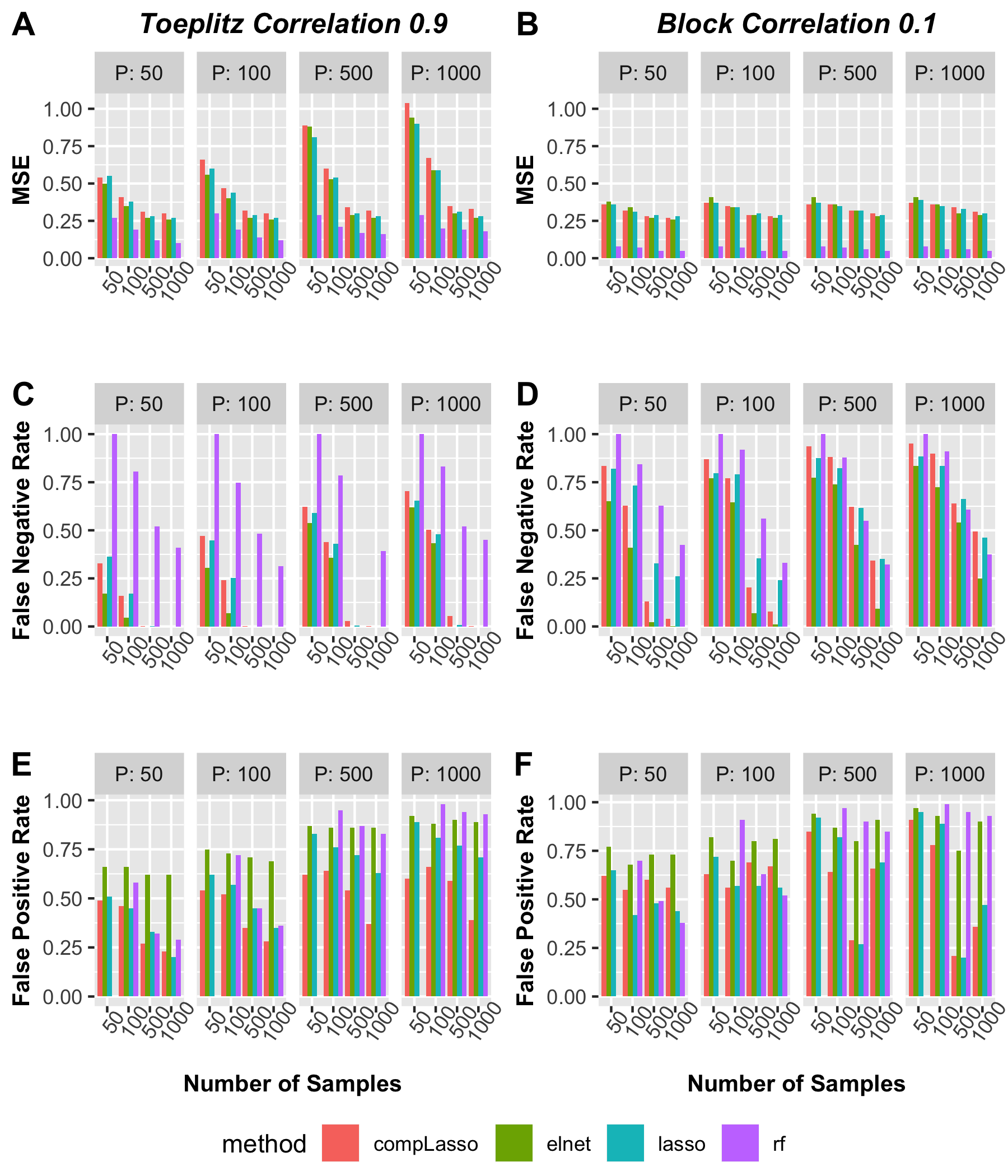}}
\caption{Method comparisons based on MSE in extreme correlation structures (Toeplitz 0.9 for A,C,E and Block 0.1 for B,D,F). Colored bars represent MSE (A-B), False Negative Rates (C-D), and False Positive Rates (E-F) corresponding to a specific number of samples (x-axis) and features ($p$) for different feature selection methods: compositional lasso (red), elastic net (green), lasso (blue) and random forests (purple). Note that false positive rates are not available for random forests when number of samples equals 50 because it chooses zero features. }
\label{f:fig3}
\end{figure}

How will results differ if we use MSE as the evaluation criterion? Using the extreme correlation settings (Toeplitz 0.9 and Block 0.1) as examples, random forests has lowest MSEs for all combinations of $p$ and $n$ (Figure \ref{f:fig3} A-B). However, Figure \ref{f:fig3} C-D unveils that random forests has highest false negative rates in all scenarios of Toeplitz 0.9 and Block 0.1, and its false negative rates can reach as high as the maximum 1, indicating that random forests fails to pick up any real signal despite its low prediction error. Moreover, Figure \ref{f:fig3} E-F show that random forests can have highest false positive rates when $p$ is as large as 500 or 1000. All these highlight the danger of choosing inappropriate feature selection method based on MSE, where the merit of high predictive power masks high errors in false positives and false negatives. On the other hand, the method with lowest false positive rates (compositional lasso) (Figure~\ref{f:fig3} E-F) was rather found to have the worst performance by MSE (Figure~\ref{f:fig3} A-B), suggesting another pitfall of missing the optimal method when using MSE as the evaluation criterion. 

The use of point estimates alone to compare feature selection methods, without incorporating variability in these estimates, could be misleading. Hence, as a next step,  we evaluate reliability of MSE and Stability across methods using a hypothesis testing framework. This is demonstrated with the cases of $n = 100$ $\&$ $p = 1000$ for Toeplitz 0.5 and Block 0.5, where compositional lasso is found to be the best feature selection method based on Stability, while random forests is the best based on MSE. We use bootstrap to construct $95\%$ confidence intervals to compare compositional lasso vs. random forests based on Stability or MSE. For each simulated data (100 in total for Toeplitz 0.5 or Block 0.5), we generate 100 bootstrapped datasets and apply feature selection methods to each bootstrapped dataset. Then for each simulated data, Stability is calculated based on the 100 subsets of selected features from the bootstrapped replicates, and the variance of Stability is measured as its variability across the 100 simulated data. Since MSE can be obtained for each simulated data without bootstrapping, we use the variability of MSE across the 100 simulated data as its variance. Based on the 95\% CI for the difference in Stability between compositional lasso and random forest methods (Table 1), we see that compositional lasso is better than random forest in terms of Stability index, and not statistically inferior to random forests in terms of MSE despite its inferior raw value. This suggests that Stability has higher precision (i.e. lower variance). Conversely, MSE has higher variance, which results in wider confidence intervals and its failure to differentiate methods. 

\begin{table}[h]
    \centering
    \label{t:table1}
    \caption{Hypothesis testing using Bootstrap to compare compositional lasso (CL) with random forests (RF) based on Stability or MSE using two simulation scenarios (*indicate statistically significant). }
    \begin{tabularx}{\textwidth}{ |X|X|X| } 
    \hline
    Example (N = 100 \& P = 1000) & Estimated mean difference (CL – RF) in Stability index with 95\% CI & Estimated mean difference (CL – RF) in MSE with 95\% CI \\ 
    \hline
    Toeplitz 0.5 & 0.22 (0.19, 0.28)*  & 0.23 (-0.62, 1.36) \\ 
    \hline
    Block 0.5 & 0.23 (0.17, 0.29)* & 0.44 (-0.27, 1.57)   \\
    \hline
    \end{tabularx}
\end{table}

\section{Experimental microbiome data applications\label{data}}

To compare the reliability of MSE and Stability in choosing feature selection methods in microbiome data applications, two experimental microbiome datasets were chosen to cover common sample types (human gut and environmental soil samples) and the scenarios of $p \approx n$ and $p >> n$ (where $p$ is the number of features and $n$ is the number of samples). 
The human gut dataset represents a cross-sectional study of 98 healthy volunteers to investigate the connections between long-term dietary patterns and gut microbiome composition \citep{wu2011linking}, and we are interested in identifying a subset of important features associated with BMI, which is a widely-used gauge of human body fat and associated with the risk of diseases. The soil dataset contains 88 samples collected from a wide array of ecosystem types in North and South~America~\citep{lauber2009pyrosequencing}, and we are interested in discovering microbial features associated with the pH gradient, as pH was reported to be a strong driver behind fluctuations in the soil microbial communities~\citep{morton2017balance}. Prior to our feature selection analysis, the same filtering procedures were applied to the microbiome count data from these two datasets, where only the microbes with a taxonomy assignment at least to genus level or lower were retained for interpretation, and microbes present in fewer than 1\% of the total samples were removed. Moreover, the count data were transformed into compositional data after replacing any zeroes by the maximum rounding error~0.5~\citep{lin2014variable}.

Comparisons of feature selection methods in these two microbiome datasets are shown in Table 2, which are consistent with simulation results, where the best method chosen by MSE or Stability in each dataset can be drastically different. Based on MSE, random forests is the best in the BMI Gut dataset, while being the worst based on Stability. Similarly, in the pH Soil dataset, random forests is the second best method according to MSE, yet the worst in terms of Stability. If we use Stability as the evaluation criterion, then Elastic Net is the best in the BMI Gut and compositional lasso is the best in the pH Soil, yet both methods would be the worst if MSE was used as the evaluation criterion. One important note is that the Stability values in these two experimental microbiome datasets are low: none of the feature selection method exceeds a stability of 0.4, indicating the challenging task of feature selection in real microbiome applications. However, this possibility of low Stability values was already reflected in our simulated scenarios of ``extreme” correlation scenarios. Another important note, which might be counter-intuitive, is that the dataset with a high $p/n$ ratio (pH Soil) has higher stabilities than the dataset with $p/n$ ratio close to 1 (i.e. similar $p$ \& $n$ values) (BMI Gut). This might be explained by the clearer microbial signals in environmental samples than in human gut samples, but it also highlights the impact of the dataset itself, whose characteristics cannot be easily summarized with the numbers of $p$ and $n$, on feature selection results. Correlation structures between features as considered in our simulations could play an important role, and there may be many other unmeasured factors involved as well. 

\begin{table}[h]
    \centering
    \label{t:table2}
    \caption{Method comparisons based on Stability Index and MSE in experimental microbiome datasets (methods ordered in terms of best MSE/Stability performance, followed with raw MSE/Stability values in parentheses). }
    \begin{tabularx}{\textwidth}{ |c|c|X|X| } 
    \hline
    Dataset & $n * p$ $(p/n)$ & MSE \newline (lower is better) & Stability \newline (higher is better) \\ 
    \hline
    BMI Gut & 98 * 87 (0.9)  & 
    Random forests (4.99) \newline
    Compositional lasso (21.59) \newline 
    Lasso (24.07) \newline
    Elastic Net (25.33) &
    Elastic Net (0.23) \newline
    Compositional lasso (0.22) \newline
    Lasso (0.14) \newline
    Random forests (0.02)\\ 
    \hline
    pH Soil & 89 * 2183 (24.5) & Elastic Net (0.23) \newline
    Random forests (0.26) \newline
    Lasso (0.34) \newline
    Compositional lasso (0.46) &
    Compositional lasso (0.39) \newline
    Lasso (0.31) \newline 
    Elastic Net (0.16) \newline
    Random forests (0.04)\\
    \hline
    \end{tabularx}
\end{table}

Apart from the comparisons based on point estimates, we can further compare MSE and Stability with hypothesis testing using nested bootstrap~\citep{wainer2018nested}. The outer bootstrap generates 100 bootstrapped replicates of the experimental microbiome datasets, and the inner bootstrap generates 100 bootstrapped dataset for each bootstrapped replicate from the outer bootstrap. Feature selections are performed on each inner bootstrapped dataset with 10-fold cross-validation after a 80:20 split of training and test sets. The variance of Stability is calculated based on the Stability values across the outer bootstrap replicates, and the variance of MSE is calculated across both inner and outer bootstrap replicates, since MSE is available for each bootstrap replicate while Stability has to be estimated based on feature selection results across multiple bootstrap replicates. Using the datasets of BMI Gut and pH Soil, Table 3 confirms with simulation results that raw value difference in MSE does not indicate statistical difference, yet difference in Stability does help to differentiate methods due to its higher precision. A comparison between the observed difference in Table 2 and the estimated mean difference from bootstrap in Table 3 further confirms this discovery. Compared to the estimated mean differences between compositional lasso and random forests based on stability (0.27 in the BMI Gut and 0.36 in the pH Soil), the observed differences (0.2 in the BMI Gut and 0.35 in the pH Soil) differ by 26\% in the BMI Gut and 3\% in the pH Soil. However,  this difference is much more drastic based on MSE. Compared to the estimated mean differences between compositional lasso and random forests based on MSE (16.6 in the BMI Gut and 0.2 in the pH Soil), the observed differences (11.8 in the BMI Gut and 0.08 in the pH Soil) have huge differences of 41\% and 160\% in each dataset respectively. Hence, Stability is consistently shown to be more appropriate than MSE in experimental data applications as in simulations. 

\begin{table}[h]
    \centering
    \label{t:table3}
    \caption{Hypothesis testing using Bootstrap to compare compositional lasso (CL) with random forests (RF) based on Stability or MSE using two experimental microbiome datasets (*indicate statistically significant).}
    \begin{tabularx}{\textwidth}{ |X|X|X| } 
    \hline
    Dataset & Estimated mean difference (CL – RF) in Stability index with 95\% CI & Estimated mean difference (CL – RF) in MSE with 95\% CI \\ 
    \hline
    BMI Gut & 0.27 (0.17, 0.34)*  & 11.8 (-2.1, 41.2) \\ 
    \hline
    pH Soil & 0.36 (0.28, 0.44)* & 0.08 (-0.28, 0.95)   \\
    \hline
    \end{tabularx}
\end{table}

\section{Discussion}
Reproducibility is imperative for any scientific discovery, but there is a growing alarm about irreproducible research results. According to a survery by Nature Publishing Group of 1,576 researchers in 2016, more than 70\% of researchers have tried and failed to reproduce another scientist’s experiments, and more than half have failed to reproduce their own experiments \citep{baker20161}. This ``reproducibility crisis” in science affects microbiome research as much as any other areas, and microbiome researchers have long struggled to make their research reproducible \citep{schloss2018identifying}. Great efforts have been made towards setting protocols and standards for microbiome data collection and processing \citep{thompson2017communal}, but more could be achieved using statistical techniques for reproducible data analysis. Microbiome research findings rely on statistical analysis of high-dimensional data, and feature selection is an indispensable component for discovering biologically relevant microbes. In this article, we focus on discovering a reproducible criterion for evaluating feature selection methods rather than developing a better feature selection method. We question the common practice of evaluating feature selection methods based on overall performance of model prediction~\citep{knights2011human}, such as Mean Squared Error (MSE), as we detect a stark contrast between prediction accuracy and reproducible feature selection. Instead, we propose to use a reproducibility criterion such as Nogueira’s Stability measurement~\citep{nogueira2017stability} for identifying the optimal feature selection method. 

In both our simulations and experimental microbiome data applications, we have shown that Stability is a preferred evaluation criterion over MSE for feature selection, because of its closer reflection of the ground truth (false positive and false negative rates) in simulations, and its better capacity to differentiate methods due to its higher precision. Hence, if the goal is to identify the underlying true biological signal, we propose to use a reproducibility criterion like Stability instead of a prediction criterion like MSE to choose feature selection algorithms for microbiome data applications. MSE is better suited for problems where prediction accuracy alone is the focus.

The strength of our work lies in the comparisons of widely used microbiome feature selection methods using extensive simulations, and experimental microbiome datasets covering various sample types and data characteristics. The comparisons are further confirmed with non-parametric hypothesis testing using bootstrap. Although Nogueira et al. were able to derive the asymptotical normal distribution of Stability~\citep{nogueira2017stability}, their independent assumption for two-sample test might not be realistic due to the fact that two feature selection methods are applied to the same dataset. Hence our non-parametric hypothesis testing is an extension of their two-sample test for Stability. However, our current usage of bootstrap, especially the nested bootstrap approach for experimental microbiome data applications, is computationally expensive; further theoretical development on hypothesis testing for reproducibility can be done to facilitate more efficient method comparisons based on Stability. Last but not least, although our paper is focused on microbiome data, we do expect the superiority of reproducibility criteria over prediction accuracy criteria in feature selection to apply in other types of datasets as well. We thus recommend that researchers use stability as an evaluation criterion while performing feature selection in order to yield reproducible results.

\section*{Acknowledgements}
We gratefully acknowledge supports from IBM Research through the AI Horizons Network, and UC San Diego AI for Healthy Living program in partnership with the UC San Diego Center for Microbiome Innovation. This work was also supported in part by CRISP, one of six centers in JUMP, a Semiconductor Research Corporation (SRC) program sponsored by DARPA. LN was partially supported by NIDDK 1R01DK110541-01A1.

\section*{Supporting Information}
The code that implements the methodology, simulations and experimental microbiome data applications is available at the Github repository https://github.com/knightlab-analyses/stability-analyses.

\label{lastpage}

\bibliographystyle{biom}
\bibliography{Bibliography}
\end{document}






\pagerange{\pageref{firstpage}--\pageref{lastpage}} 


\doi{10.1111/j.1541-0420.2005.00454.x}


\label{firstpage}



\maketitle


%

\beginsupplement
\section*{Supplementary Figures}

\begin{suppfigure}[h]
 \centerline{\includegraphics[width=5in]{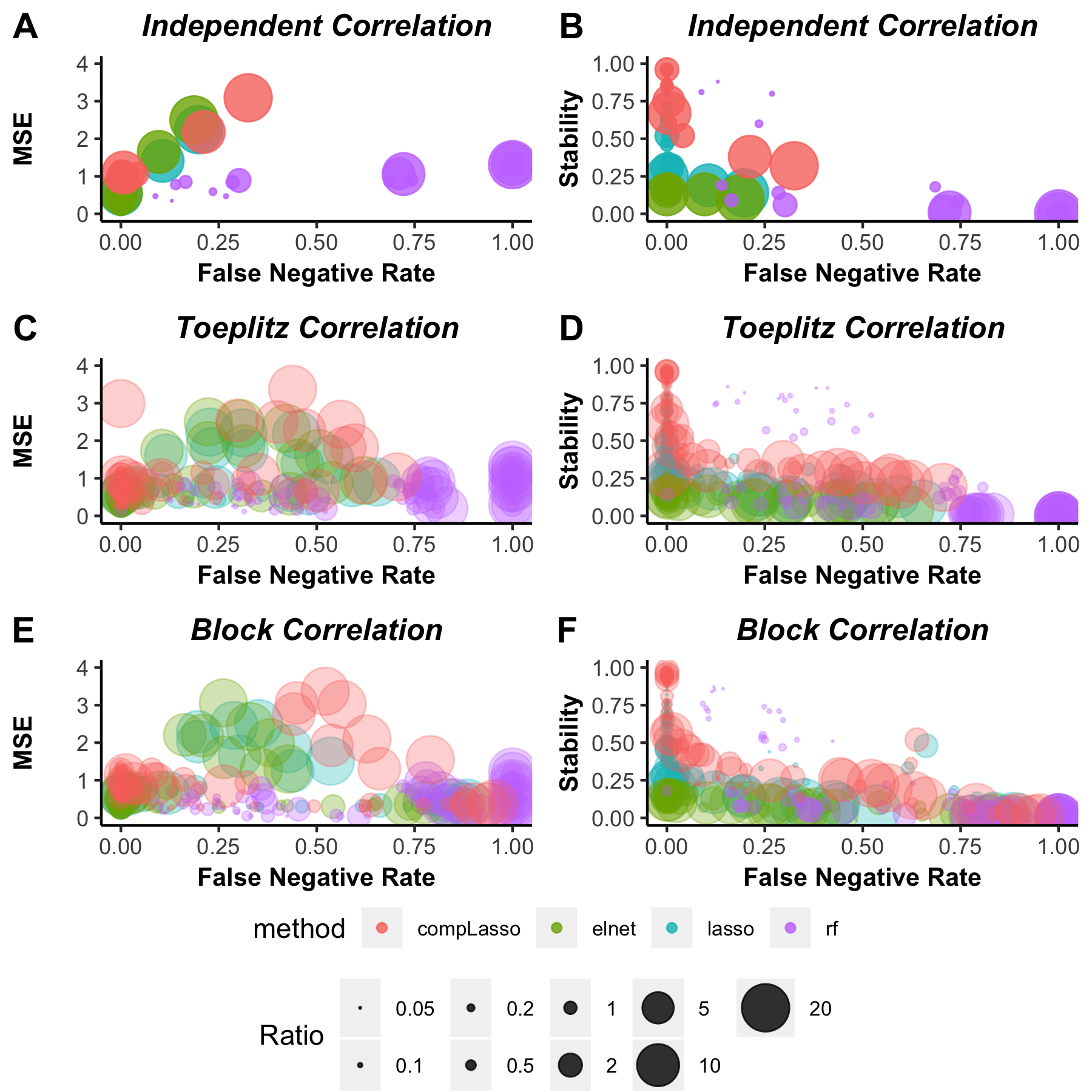}}
\caption{Compare the relationship between MSE and False Negative Rate vs. Stability and False Negative Rate in three correlation structures. Colored dots represent values from different feature selection methods: compositional Lasso (red), Elastic Net (green), Lasso (blue) and random forests (purple). Size of dots indicate features-to-sample size ratio  $p/n$.}
\label{f:figS1}
\end{suppfigure}

\begin{suppfigure}
 \centerline{\includegraphics[width=5in]{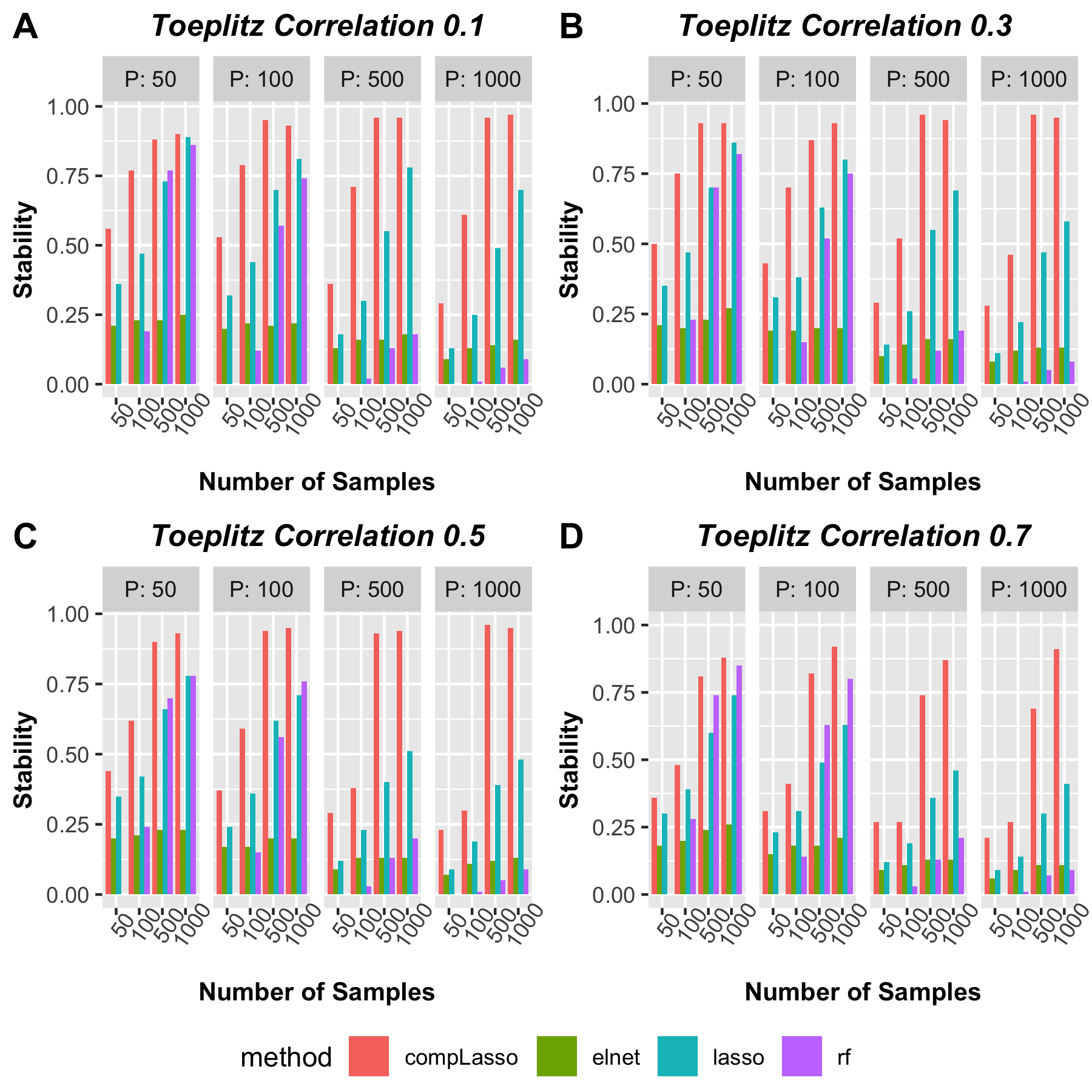}}
\caption{Method comparisons based on Stability in easier Toeplitz correlation structures from 0.1 to 0.7. Colored bars represent Stability values corresponding to specific number of samples (x-axis) and number of features (p) for different feature selection methods: compositional Lasso (red), Elastic Net (green), Lasso (blue) and random forests (purple). Compositional Lasso has highest stability in all cases across all correlation strength. Note that Stability equals to zero when no features were selected by methods (e.g. random forests chooses nothing when the number of samples equals 50).}
\label{f:figS2}
\end{suppfigure}

\begin{suppfigure}
 \centerline{\includegraphics[width=5in]{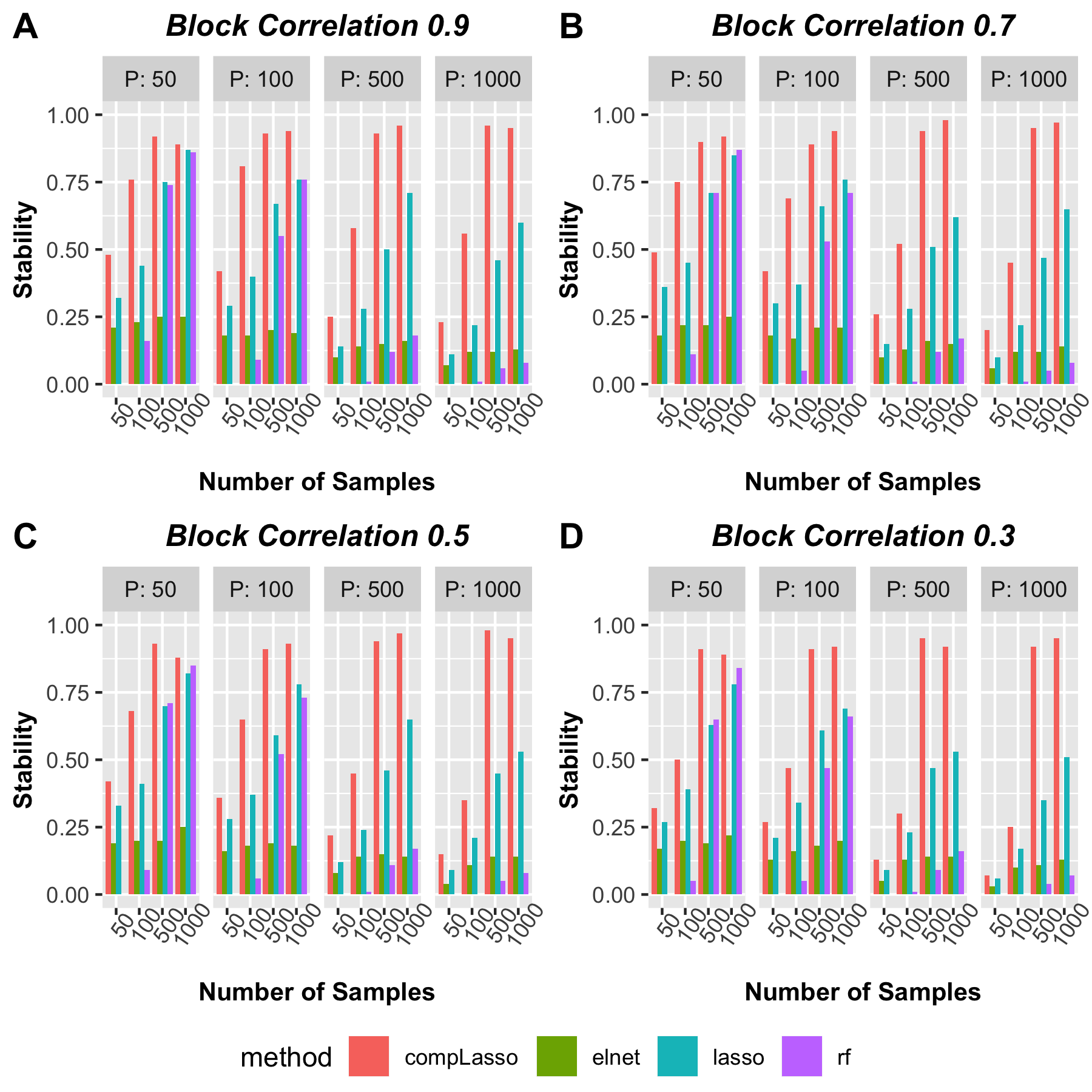}}
\caption{Method comparisons based on Stability in easier Block correlation structures from 0.9 to 0.3. Colored bars represent Stability values corresponding to specific number of samples (x-axis) and number of features (p) for different feature selection methods: compositional Lasso (red), Elastic Net (green), Lasso (blue) and random forests (purple). Compositional Lasso has highest stability in all cases across all correlation strength. Note that Stability equals to zero when no features were selected by methods (e.g. random forests chooses nothing when the number of samples equals 50).}
\label{f:figS3}
\end{suppfigure}

\label{lastpage}